\documentclass[oupdraft]{biostatistics}
\usepackage[utf8]{inputenc}
\usepackage{amsmath}
\DeclareMathOperator*{\argmax}{arg\,max}

\newcommand\numberthis{\addtocounter{equation}{1}\tag{\theequation}}
\usepackage{amssymb}
\usepackage{float}
\usepackage{mathtools}
\usepackage{graphicx}
\usepackage{multirow}
\usepackage{colortbl}
\usepackage{comment}
\usepackage{adjustbox}
\usepackage{pdflscape}
\usepackage{rotating}
\floatstyle{plaintop}
\usepackage{xcolor}
\usepackage{fancyhdr}
\usepackage{url}

\title{A meta-inference framework to integrate multiple external models into a current study}
\date{\today}

\begin{document}

\author{Tian Gu$^\ast$, Jeremy M.G. Taylor and Bhramar Mukherjee\\[4pt]
\textit{Department of Biostatistics, University of Michigan, Ann Arbor, MI 48109, USA}
\\[2pt]
{gutian@umich.edu}}

\markboth%
{T. Gu \& J.M.G. Taylor \& B. Mukherjee}
{A meta-inference framework}

\maketitle

\footnotetext{To whom correspondence should be addressed.}

\begin{abstract}
{It is becoming increasingly common for researchers to consider incorporating external information from large studies to improve the accuracy of statistical inference instead of relying on a modestly sized dataset collected internally. With some new predictors only available internally, we aim to build improved regression models based on individual-level data from an “internal” study while incorporating summary-level information from “external” models. We propose a meta-analysis framework along with two weighted estimators as the composite of empirical Bayes estimators, which combines the estimates from the different external models. The proposed framework is flexible and robust in the ways that (i) it is capable of incorporating external models that use a slightly different set of covariates; (ii) it is able to identify the most relevant external information and diminish the influence of information that is less compatible with the internal data; and (iii) it nicely balances the bias-variance trade-off while preserving the most efficiency gain. The proposed estimators are more efficient than the naïve analysis of the internal data and other naïve combinations of external estimators.}{Data integration; Prediction models; Empirical Bayes; Meta-analysis.}
\end{abstract}

\section{Introduction}
In the era of big data, it is becoming increasingly common for researchers to consider incorporating external information from large datasets or studies to improve the accuracy of statistical inference instead of relying on a modestly sized dataset collected internally. For example, borrowing strength from historical control data to leverage the treatment effect in small-sample clinical trials \citep{Viele2014, Dejardin2018, Li2020}, combining separate probability samples \citep{Bycroft2010, YangKim2020} and incorporating external data sources for improved causal inference \citep{YangDing2020}. However, challenges exist, such as data sharing, storage, and privacy issues to access publicly available individual-level large data, so often only the summary information is reported. Examples of such data sources include publications and online risk calculators. Therefore, general frameworks that integrate the individual-level data and the summary-level external information are particularly needed. As a motivating example, it is common in predictive modeling that researchers want to include new predictors to update the traditional risk models in clinical biomedicine, such as adding genetic risk variants and mammographic density to the breast cancer risk calculator \cite{Gail1989}. Since the new variables are often only available in a moderate-sized study, it is natural to consider incorporating the external model information for improved inference. To effectively utilize the external information, the external population needs to share some distributional features with the internal population, which is often referred to as transportability \citep{Bareinboim2013} in causal inference.

Recent studies on incorporating external summary information into the regression estimation include both frequentist \citep{Qin2000, Chatterjee2016, Han2019, Chatterjee2019, Gu2019, Zhang2020} and Bayesian approaches \citep{Cheng2018, Cheng2019}. Several methods were built upon the work of \citet{Qin2000} and \citet{Chatterjee2016}, who described a constrained semi-parametric maximum likelihood (CSPML) method by converting the external summary-level information into a constraint and then maximizing the internal data likelihood subject to this constraint. However, the CSPML method requires the joint distribution of (Y, X, B) to be the same in the internal and the external population, a strong assumption, which although unverifiable, we expect would be frequently violated, and can cause bias when violated. \citet{Estes2017} later proposed a matrix-weighted average remedy by constructing an empirical Bayes (EB) estimator that can reduce the potential bias. As an extension and adaption of \citet{Estes2017}, we propose a meta-inference framework using a composite of empirical Bayes estimators to accommodate the situation where multiple external prediction models are available to help improve the inference of the current study.

We consider the situation in which there are K external studies (K $\ge$ 2), each of which developed a prediction model for the same outcome. The parameter estimates of the external models are known, but the individual level data are not available. The goal is to develop a prediction model that uses all the possible covariates, using data from an internal study and the parameter estimates from the external models. The parameters of this model are the quantities of interest. Each of the external studies may use a slightly different set of covariates but the internal data are assumed to contain all available covariates, as well as the new biomarkers that are not included in any of the external models. We propose a meta-inference framework using an empirical Bayes estimation approach, which first separately incorporates the different summary information from each external study into the internal study, and then takes a weighted average of the resulting estimators to give a final overall estimate of the parameters of interest. We show that the proposed final estimators are more efficient than the simple analysis of the internal data, as well as outperform the estimators that integrate the information from a single external model.

\section{Models and Methods}
\subsection{General Description of the Problem} \label{ch3_notation}
Let $\rm Y$ denote the outcome of interest, which can be either continuous or binary. $\rm \mathbf{X}$ is a set of p standard variables and let $\rm B$ denote a new biomarker. Our target of interest is the mean structure of $\rm Y|\mathbf{X}, B$:
  \begin{equation} \label{Ch3_general Y|XB}
        g(\rm E(Y|\mathbf{X},B))\\
        = \boldsymbol{\rm X \gamma}_X + B \gamma_B = \gamma_{X_0} + \gamma_{X_1} X_1 + ... + \gamma_{X_p} X_p + \gamma_B B,
    \end{equation}
where $g$ is the known link function. We assume that a small dataset of size n with variables $\rm Y, \mathbf{X}$ and a new covariate $\rm B$ is available to us for building model \ref{ch3_notation}. For each external study k, k $\in \rm \{2,...,K\}$, a prediction model for the same outcome $\rm Y$ has been built using predictors $\rm \mathbf{X}_k$, a subset of the internal $\mathbf{X}$. Each external model may use slightly different predictors to predict $\rm Y$:
    \begin{equation*}
        g(\rm E(Y|\mathbf{X}_k))\\
        = \boldsymbol{\rm X}_k \boldsymbol{\rm \beta}_k = \rm \beta_0 + \beta_1 X_1 + \dots + \beta_{p_k} X_{p_k},
    \end{equation*}
where $\rm p_k \subseteq p$ is the dimension of $\rm \mathbf{X}_k$. We assume that the distribution of $\rm Y|\mathbf{X}, B$ is correctly specified, but the external $\rm Y|\mathbf{X}_k$ distributions need not be.

We assume K large, well-characterized previous studies from the external populations describe the provided information on the calculated distribution of $\rm Y|\mathbf{X}_k$. These information will come in the forms of estimated model parameters $\rm \hat{\boldsymbol{\beta}}_k$. The goal is to develop a framework, in which we can utilize all K external $\rm \hat{\boldsymbol{\beta}}_k$'s to improve the estimation efficiency of the internal study.

We introduce some of the important notation that will frequently appear in later sections:

\begin{itemize}
    \item $\rm f_{\beta}(Y|X_k)$: the study-specific distribution of the $\rm k^{th}$ external model $\rm Y|\mathbf{X}_k$;
    \item $\rm f_{\gamma}(Y|X,B)$: distribution of the target model $\rm Y|\mathbf{X},B$;
    \item $\rm \hat{\boldsymbol{\gamma}}_I$: the unconstrained estimator by using the internal data only;
    \item $\rm \hat{\boldsymbol{\gamma}}_{CML}$: the constrained semi-parametric maximum likelihood estimator (CSPML) proposed by \citet{Chatterjee2016};
    \item $\rm \hat{\boldsymbol{\gamma}}_{EB}$: the empirical Bayes (EB) estimator proposed by \citet{Estes2017}.
\end{itemize}

\subsection{Two Existing Estimators} \label{ch3_Chatterjee and Estes}
The proposed method was developed on the foundation of two existing methods, the CSPML approach \citep{Chatterjee2016} and the EB approach \citep{Estes2017}. The CSPML estimator considered the same problem described here in a special case where K=1; and EB estimator applied the empirical Bayes method to CSPML, calibrating the potential bias due to non-transportability. Therefore, it is necessary to introduce these two core methods first before considering the $\rm K>1$ situation. 

In the CSPML, the proposed estimator $\rm \hat{\boldsymbol{\gamma}}_{CML}$ incorporates the external regression coefficients to calibrate the current regression model. Denote $\rm U_{\beta}(Y|X)$ as the score function of the external $\rm Y|X; \beta$ model. It converts the external model parameter $\rm \hat{\boldsymbol{\beta}}$ to a constraint by connecting the external score function with the target distribution $\rm f_{\gamma}(Y|X,B)$:
\begin{align*} 
    \rm 0 = \rm E_{Y,X,B}[U_{\beta}(Y|X)]
    &= \rm E_{X,B}\{E_{Y|X,B}[U_{\beta}(Y|X) |X,B]\}\\
    &=\rm \int_{X,B}{\int_{Y|X,B}  U_{\beta}(Y|X) f_{\gamma}(Y|X,B)dY}dF(X,B)\\
    &= \rm \sum_{i=1}^n \int_{Y|X,B} U_{\beta}(Y|X) f_{\gamma}(Y|X,B)dY p_i,
\end{align*} where $\rm dF(X,B)$ is the empirical probability distribution $\rm p_i=Pr(X=X_i,B=B_i)$ for the internal observations and $\rm \sum_{i=1}^n p_i=1$. Then $\rm \hat{\boldsymbol{\gamma}}_{CML}$ was obtained using Lagrange multipliers by solving the following Lagrangian function:
\begin{equation*}
        \rm \boldsymbol{\hat{\gamma}}_{CML} = \argmax_{\rm \gamma, p_i} \{\prod_{i=1}^{n} \rm f_{\gamma}(Y_i|X_i,B_i)p_i + \rm \lambda_1 (\sum\limits_{i=1}^n p_i - 1)+ \rm \lambda_2 \sum\limits_{i=1}^n \int U_{\beta}(Y|X) f_{\gamma}(Y|X,B) dY p_i  \}
\end{equation*}
\citet{Chatterjee2016} provided the asymptotic variance of $\rm \hat{\boldsymbol{\gamma}}_{CML}$, showing the efficiency gain of $\rm \hat{\boldsymbol{\gamma}}_{CML}$ compared to $\rm \hat{\boldsymbol{\gamma}}_I$. 

\citet{Estes2017} showed that in the CSPML approach, the strict assumption of the identical joint distribution of (Y,X,B) between the internal and the external population (also known as the full transportable assumption of the joint distribution of [Y,X,B]) is hard to satisfy in reality, and ignoring it can lead to substantial bias. Assuming the target conditional distribution ($\rm Y|X, B; \gamma$) is correctly specified in the internal study and the underlying true parameter $\rm \gamma$ follows a stochastic framework, \citet{Estes2017} proposed an empirical Bayes (EB) estimator $\rm \hat{\boldsymbol{\gamma}}_{EB}$, which can be viewed as a matrix-weighted average of the internal estimate $\rm \hat{\boldsymbol{\gamma}}_I$ and the CSPML estimator $\rm \hat{\boldsymbol{\gamma}}_{CML}$. The EB estimator uses the difference $\rm \hat{\boldsymbol{\gamma}}_I - \hat{\boldsymbol{\gamma}}_{CML}$ to measure the distributional similarity of the joint distribution (Y, X, B) between the internal and the external population, and will down-weight $\rm \hat{\boldsymbol{\gamma}}_{CML}$ if the lack of full transportability leads to a poor estimate. Therefore, the EB estimator is robust to departures from full transportability assumption in specific external populations. The features and assumptions for the CSPML, the EB and the proposed estimators are summarized in Table A1 in Appendix A of the Supplementary Material.

Specifically, the EB approach first posits a stochastic framework connecting the internal estimator $\rm \hat{\boldsymbol{\gamma}}_I$ and the underlying true parameter $\rm \boldsymbol\gamma \sim N(\boldsymbol\gamma_0, \mathbf{A})$ for some covariance matrix $\rm \mathbf{A}$. Since 
$\begin{cases}
    \rm \boldsymbol\gamma \sim N(\boldsymbol\gamma_0, \mathbf{A})\\
    \rm \hat{\boldsymbol\gamma}_I|\boldsymbol\gamma~N(\boldsymbol\gamma, \mathbf{\Sigma})
\end{cases}$, the posterior Bayes estimate of $\rm \boldsymbol\gamma$ equals to $\rm \mathbf{A}(\boldsymbol\Sigma+\mathbf{A})^{-1} \hat{\boldsymbol\gamma}_I+\boldsymbol\Sigma(\boldsymbol\Sigma+\mathbf{A})^{-1}\boldsymbol\gamma_0$. Replacing $\rm \boldsymbol\gamma_0$ with the CSPML estimator $\rm \hat{\boldsymbol\gamma}_{CML}$ and empirically estimating $\rm \mathbf{A}$ and $\rm \boldsymbol\Sigma$, we obtain the EB estimator $\rm \hat{\boldsymbol\gamma}_{EB} = \hat{\mathbf{A}}(\hat{\boldsymbol\Sigma}+\hat{\mathbf{A}})^{-1} \hat{\boldsymbol\gamma}_I+\hat{\boldsymbol\Sigma}(\hat{\boldsymbol\Sigma}+\hat{\mathbf{A}})^{-1}\hat{\boldsymbol\gamma}_{CML} \overset{\text{def}}{=} \rm \hat{\mathbf{W}} \boldsymbol{\hat{\gamma}}_I  + (I-\hat{\mathbf{W}}) \boldsymbol{\hat{\gamma}}_{CML}$, where $\rm \hat{\mathbf{A}} = (\boldsymbol{\hat{
\gamma}}_I - \boldsymbol{\hat{\gamma}}_{CML})(\boldsymbol{\hat{
\gamma}}_I - \boldsymbol{\hat{\gamma}}_{CML})^T$ quantifies the difference between $\rm \hat{\boldsymbol{\gamma}}_I$ and $\rm \boldsymbol{\hat{\gamma}}_{CML}$, $\rm \hat{\boldsymbol\Sigma}$ is the MLE of the variance of $\rm \boldsymbol{\hat{
\gamma}}_I$, and $\rm \hat{\mathbf{W}} = \hat{\mathbf{A}}(\hat{\boldsymbol\Sigma}+\hat{\mathbf{A}})^{-1}$ is the empirical weights. Therefore, the EB estimator can be viewed as a matrix generalization of a weighted average of vectors $\rm \boldsymbol{\hat{\gamma}}_I$ and $\rm \boldsymbol{\hat{\gamma}}_{CML}$, of which the empirical weights will shrink the EB estimator towards $\rm \boldsymbol{\hat{\gamma}}_I$ when $\rm \boldsymbol{\hat{\gamma}}_{CML}$ is biased. The EB estimator's shrinkage effect limits the impact of external model information that is not compatible with the internal data and thus protects against the severe bias. Furthermore, when the joint distribution of (Y,X,B) is similar in the two populations, more precision will be gained by incorporating the external model.

\subsection{Proposed Meta-Framework for Inference} \label{proposed framework}
We build upon the empirical Bayes work by \citet{Estes2017} and generalize it to accommodate the situation where we can combine multiple external model estimates into the internal study. Our proposed meta-inference framework allows each established model to have different dimension. The framework consists of generating an EB estimator for each of the external $\rm \hat{\boldsymbol{\beta}}_k$ from the fitted regression $\rm Y|\mathbf{X}_k$, and then constructing the final estimates of the target through a weighted average, considering the correlation structure among all the EB estimators.

The proposed framework contains two steps:
\begin{itemize}
    \item Step 1: For each of the K external model estimates $\rm \hat{\boldsymbol{\beta}}_k$, first apply the CSPML method \citep{Chatterjee2016}, and then apply the EB method \citep{Estes2017}:
    \begin{equation*}
        \text{Internal data + External } \hat{\boldsymbol{\beta}}_k \xrightarrow{\text{CSPML}} \rm \hat{\boldsymbol{\gamma}}_{CML_k} \xrightarrow{\text{EB}} \rm \hat{\boldsymbol{\gamma}}_{EB_k}=\rm \hat{W}_k \boldsymbol{\hat{\gamma}}_I  + (I-\hat{W}_k) \boldsymbol{\hat{\gamma}}_{CML_k},
    \end{equation*} after which we obtain a total of K $\rm \hat{\boldsymbol{\gamma}}_{EB}$'s.
    
    \item Step 2: We propose two estimators of $\rm \boldsymbol{\gamma}$ to be the weighted average of $\rm \hat{\boldsymbol{\gamma}}_{EB}$'s, of the form
    $\rm \sum_{k=1}^K w_{k} \hat{\boldsymbol{\gamma}}_{EB_k}$,
    in which each element of the final estimate is the weighted average of the $\rm K$ separate estimates for that element.
    One composite estimator is called the optimal covariance weighted estimator (OCWE) and another is called the selective coefficient learner (SC-Learner).
\end{itemize}

In step 1, we separately integrate each of the external $\rm \hat{\boldsymbol{\beta}}$'s with the internal data, and the EB method accounts for the potential bias caused by the heterogeneity of the internal and that specific external population. This first step also unifies the disparate dimensions of the external models to be the same as the target model \ref{Ch3_general Y|XB}, and improves the efficiency of parameter estimates for those covariates that were used in the external models. 

In step 2, the challenge is to combine K correlated vectors of EB estimators while maximizing the efficiency gain of the overall prediction. The simplest, yet not the most attractive, solution is to naively average $\rm \hat{\boldsymbol{\gamma}}_{EB^{'}s}$, i.e. $\rm  \frac{1}{K}\sum_{k=1}^K \rm \boldsymbol{\hat{\gamma}}_{EB_k}$. Better weighting approaches take into account the variance and/or correlation among $\rm \hat{\boldsymbol{\gamma}}_{EB^{'}s}$. One option is to use the inverse of the prediction variances as weights, i.e. $\rm w_k=\frac{1/\sum_{i=1}^n \hat{V}ar[(\rm \boldsymbol{\rm X_i},B_i) \hat{\boldsymbol{\gamma}}_{EB_k}]} {\sum_{k=1}^K 1/\sum_{i=1}^n \hat{V}ar[(\rm \boldsymbol{\rm X_i},B_i) \hat{\boldsymbol{\gamma}}_{EB_k}]}$, with the same $\rm w_k$ used for all elements of $\boldsymbol{\gamma}$. This method incorporates the within-estimator variance while ignoring the between-estimator covariance (i.e. ignoring the fact that the $\rm \hat{\boldsymbol{\gamma}}_{EB}$'s are not independent). Other popular design criterions that seek the optimal $\rm \hat{\boldsymbol{\gamma}}$ to minimize the variance-covariance matrix of $\rm \boldsymbol{\gamma}$ include D-optimality that minimizes the determinant of the matrix, A-optimality that minimizes the trace of the matrix, I-optimality (also known as V- or IV- or Q-optimality) that minimize the average prediction variance and G-optimality that minimizes the maximum prediction variance \citep{D-optimal}. Since all criterions had similar performance in this study, we consider an adaptive version of I-optimality which seeks $\rm \hat{\boldsymbol{\gamma}}$ that minimizes the average variance of the predicted estimator $\rm \frac{1}{n}\sum_{i=1}^n \hat{V}ar[(\rm \boldsymbol{\rm X_i},B_i) \hat{\boldsymbol{\gamma}}]$. We propose two weighted estimators that accounts for both within and between variances among $\rm \hat{\boldsymbol{\gamma}}_{EB}$'s:
\begin{enumerate}
    \item \textbf{The optimal covariance weighted estimator (OCWE)}: 
    OCWE views each $\rm \hat{\boldsymbol{\gamma}}_{EB}$ as a whole and provides the same weight $\rm w_k$ to each covariate coefficient within $\rm \hat{\boldsymbol{\gamma}}_{EB_k}$ that minimizes the overall estimated prediction variance, i.e.
    \[
    \rm \hat{\boldsymbol{\gamma}}_{OCWE} =
    \underset{\mathbf{w}}{\mathrm{argmin}}\sum\limits_{i=1}^n \rm \hat{V}ar[(\rm \boldsymbol{\rm X_i},B_i) \hat{\boldsymbol{\gamma}}(\mathbf{w})],
    \]
    where $\rm \boldsymbol{\rm \hat{\gamma}}(\mathbf{w}) = \rm  \sum\limits_{k=1}^K w_k \rm \boldsymbol{\hat{\gamma}}_{EB_k}$ and $\boldsymbol{\rm \mathbf{w}} = (\rm w_1, \dots, w_K)^T$ denotes the positive weights that add up to one. 
    
    \item \textbf{The selective coefficient learner (SC-Learner)}: 
    Instead of seeking a fixed weight for each $\rm \hat{\boldsymbol{\gamma}}_{EB}$ as in OCWE, SC-Learner attempts to find a set of weights separately for each covariate coefficient (from intercept $\rm \hat{\gamma}_0$ to slopes $\rm \hat{\gamma}_{X_1},...\hat{\gamma}_{X_p},\hat{\gamma}_B$) that minimize the corresponding variance, and thus each coefficient in one $\rm \hat{\boldsymbol{\gamma}}_{EB}$ can have different weights. Let $\rm E_j$ denotes the index set of the external models that included the predictor $\rm X_j$. For each predictor $\rm X_j$ (j $\in$ 0,...,p), SC-Learner first selects $\rm \hat{\gamma}_{X_j^k}$ from $\rm \hat{\boldsymbol{\gamma}}_{EB_k}$ which used $\rm X_j$ as a predictor in the external model, and then uses the inverse variance as the weight $\rm w_{kj} = \frac{1/\hat{V}ar(\boldsymbol{\hat{\gamma}}_{X_j^k})}{\sum_{k\in E_j} 1/\hat{V}ar(\boldsymbol{\hat{\gamma}}_{X_j^k})}$. The final estimate of each $\rm \gamma_{X_j}$ is an inverse variance-weighted estimator using selective coefficients from $\rm \hat{\boldsymbol{\gamma}}_{EB}$'s:
    $\rm \hat{\gamma}_{X_j}^{*} = \rm \sum_{k\in E_j} w_{kj}\boldsymbol{\hat{\gamma}}_{X_j^k}$.
    We will use $\rm \hat{\gamma}_{B}$ from the direct regression $\rm \hat{\boldsymbol{\gamma}}_I$ as the final estimate for the $\rm B$ variable, since $\rm B$ is only available from the internal data and no external models have used $\rm B$ as predictors. Thus, 
    \begin{equation*}
        \rm \hat{\boldsymbol{\gamma}}_{SC-Learner} =
            [\rm \hat{\gamma}_{X_0}^{*},\hat{\gamma}_{X_1}^{*},..., \hat{\gamma}_{X_p}^{*}, \hat{\gamma}_B]^T
    \end{equation*} 
    To illustrate this method, consider a hypothetical example with three external models--Model 1 $\rm Y|X_1, X_2, X_3$, Model 2 $\rm Y|X_1, X_2$, and Model 3 $\rm Y|X_1, X_3$ available to build the target model $\rm Y|X_1, X_2, X_3, B$ together with the internal data. When considering the final estimated coefficient of $\rm X_3$, $\rm \hat{\gamma}_{X_3^1}$ and $\rm \hat{\gamma}_{X_3^3}$ from the external models 1 and 3 will be used, while $\rm \hat{\gamma}_{X_3^2}$ that did not add extra information to $\rm X_3$ will be excluded. 
\end{enumerate}
In both proposed estimators, we use the asymptotic variance-covariance structure derived from the large sample theory to capture the correlation among $\rm \hat{\boldsymbol{\gamma}}_{EB^{'}s}$, which will be discussed in detail in Section \ref{Asymptotic_theory}.

\subsection{Asymptotic normality and large sample results} \label{Asymptotic_theory}
The following proposition extends the asymptotic normality
of the CSPML estimator \citep{Chatterjee2016} to higher dimension, as well as showing the correlation structure between $\rm \hat{\boldsymbol{\gamma}}_{CML}$ and $\rm \hat{\boldsymbol{\gamma}}_I$.

$\textit{Proposition 1.}$ Let $\rm \boldsymbol{\hat{\eta}} = \rm (\hat{\boldsymbol{\gamma}}_{CML_1}^T,\dots,\hat{\boldsymbol{\gamma}}_{CML_K}^T, \hat{\boldsymbol{\gamma}}_I^T)^T$, and $\rm \boldsymbol{\eta}_0 = \rm (\boldsymbol{\gamma}_0^T,\dots,\boldsymbol{\gamma}_0^T, \boldsymbol{\gamma}_0^T)^T$ with $\rm \boldsymbol{\gamma}_0$ the true value of $\rm \boldsymbol{\gamma}_{CML}$ and $\rm \boldsymbol{\gamma}_I$. Under regularity conditions described in \cite{Chatterjee2016}, as the internal sample size n $\rightarrow \infty$, $\sqrt{n}(\rm \boldsymbol{\hat{\eta}} - \rm \boldsymbol{\eta}_0)$ converges in distribution to a normal distribution with zero-mean and covariance matrix given by
\begin{equation} \label{ch3_Var(CML)1}
\footnotesize
    \begin{pmatrix}
        \rm (B+C_1^TL_{11}^{-1}C_1)^{-1} & \dots  & \dots & \rm \dots & \dots & \rm (B+C_1^TL_{11}^{-1}C_1)^{-1} \\
        & \ddots &  &\vdots & & \vdots\\
        & \rm (B+C_j^TL_{jj}^{-1}C_j)^{-1} & \dots & \rm Cov(\hat{\boldsymbol{\gamma}}_{CML_j}, \hat{\boldsymbol{\gamma}}_{CML_k}) & \dots & \rm (B+C_j^TL_{jj}^{-1}C_j)^{-1}\\
        & & \ddots & \vdots & & \vdots\\
        & & &\rm (B+C_K^TL_{KK}^{-1}C_K)^{-1} & \dots & \rm (B+C_K^TL_{KK}^{-1}C_K)^{-1}\\
        & & & & & \rm B^{-1}
    \end{pmatrix},
\end{equation}
where $\rm B = -E\{\frac{\partial^2}{\partial \gamma^T\gamma}logf_{\gamma}(Y|X,B)\}$, 
$\rm u_{\gamma}(\beta_j)=E_{Y|X,B}\{\frac{\partial}{\partial \beta_j}logf_{\beta_j}(Y|X)\}$, 
$\rm C_j = E\{\frac{\partial}{\partial \gamma}u_{\gamma}(\beta_j)\}$, 
$\rm L_{jk} = E\{u_{\gamma}^T(\beta_j) u_{\gamma}(\beta_k))\}$, $\rm Cov(\hat{\boldsymbol{\gamma}}_{CML_j}, \hat{\boldsymbol{\gamma}}_{CML_k})=(B+C_j^TL_{jj}^{-1}C_j)^{-1}(B+C_{j}L_{jj}^{-1}L_{jk}L_{kk}^{-1}C_k)(B+C_k^TL_{kk}^{-1}C_k)^{-1}$, and
$\rm j, k \in \{1,\dots,K\}$.

As shown in \it{Proposition 1}, \rm as well as in \cite{Chatterjee2016}, the asymptotic variance of $\rm \hat{\boldsymbol{\gamma}}_{CML}$ will decrease from $\rm B^{-1}$ to $\rm (B+C_k^TL_{kk}^{-1}C_{k})^{-1}$ after incorporating the $\rm k^{th}$ external model information. We further show additional two conclusions: (i) the covariance between $\rm \hat{\boldsymbol{\gamma}}_{CML_k}$ and $\rm \hat{\boldsymbol{\gamma}}_I$ is equivalent to the variance of $\rm \hat{\boldsymbol{\gamma}}_{CML_k}$, i.e. $\rm Cov(\hat{\boldsymbol{\gamma}}_{CML_k}, \hat{\boldsymbol{\gamma}}_I) = Var(\hat{\boldsymbol{\gamma}}_{CML_k})$; (ii) the covariance between any two $\rm \hat{\boldsymbol{\gamma}}_{CML_j}$ and $\rm \hat{\boldsymbol{\gamma}}_{CML_k}$ equals to ($\rm B+C_jL_{jj}^{-1}L_{jk}L_{kk}^{-1}C_k$) multiplied by their variances, i.e. $\rm Cov(\hat{\boldsymbol{\gamma}}_{CML_j}, \hat{\boldsymbol{\gamma}}_{CML_k}) = (B+C_j^TL_{jj}^{-1}C_j)^{-1}(B+C_jL_{jj}^{-1}L_{jk}L_{kk}^{-1}C_k)(B+C_k^TL_{kk}^{-1}C_k)^{-1}$. In Appendix B of the Supplementary Material, we show the extension of \it{Proposition 1} \rm when the uncertainty of the external $\rm \hat{\boldsymbol{\beta}}$ is known. As expected, this modification makes a difference only when the uncertainty is large.

$\textit{Proposition 2.}$ Let $\rm Z \equiv \hat{\boldsymbol{\gamma}_I} - \hat{\boldsymbol{\gamma}}_{CML}$ and $\rm \hat{V}_I \equiv \hat{Var}(\hat{\boldsymbol{\gamma}}_I)$. We can re-parameterize $\rm \hat{\boldsymbol{\gamma}}_{EB}$ as a function of $\rm Z$ and $\rm \hat{\boldsymbol{\gamma}}_{CML}$:
\[
\rm \hat{\boldsymbol{\gamma}}_{EB} = \rm \rm \hat{\boldsymbol{\gamma}}_{CML} + Z(1-\frac{1}{1+Z^T\hat{V}_I^{-1}Z}),
\]
where $\rm Z^T\hat{V}_I^{-1}Z$ is a scalar. Equivalently, $\rm \hat{\boldsymbol{\gamma}}_{EB}$ can be written as a function of $\rm Z$ and $\rm \hat{\boldsymbol{\gamma}}_I$, i.e. $\rm \hat{\boldsymbol{\gamma}}_{EB} = \rm \hat{\boldsymbol{\gamma}}_I - Z \frac{1}{1+Z^T\hat{V}_I^{-1}Z}$. The proof is listed in Appendix C of the Supplementary Material.

In the proposed method, we use the asymptotic variance-covariance structure derived from \it{Proposition 1} \rm and \it{2} \rm to capture the correlation among $\rm \hat{\boldsymbol{\gamma}}_{EB^{'}s}$. Under the assumption that the external population is representative of the target population of interest, the mean of $\rm Z$ converges to zero. In Appendix C of the Supplementary Material, we show that $\rm Z$ and $\rm \hat{\boldsymbol{\gamma}}_{CML}$ are independently normal-distributed with closed-form mean and variance, from which it is easy to simulate many values of  $\rm Z$ and $\rm \hat{\boldsymbol{\gamma}}_{CML}$. Therefore, according to \it{Proposition 2} \rm and denote $\rm f(Z)$ as $\rm Z(1-\frac{1}{1+Z^T\hat{V}_I^{-1}Z})$, we can easily obtain the numeric values of $\rm Var(\hat{\boldsymbol{\gamma}}_{EB_k})$ through the equation $\rm Var(\hat{\boldsymbol{\gamma}}_{EB_k})$ = $\rm Var(\hat{\boldsymbol{\gamma}}_{CML_k})$+$\rm Var[f(Z)]$+2$\rm  Cov[\hat{\boldsymbol{\gamma}}_{CML_k}, f(Z)]$, using the simulated values of $\rm Z$ and $\rm \hat{\boldsymbol{\gamma}}_{CML}$. A similar idea can be applied to obtain $\rm Cov(\hat{\boldsymbol{\gamma}}_{EB_j}, \hat{\boldsymbol{\gamma}}_{EB_k})$. In the simulation results, we show that when $\rm \hat{\boldsymbol{\gamma}}_{CML}$ differs from $\rm \hat{\boldsymbol{\gamma}}_I$, i.e. $\rm Z \nrightarrow 0$, the impact on the variance calculation is moderately acceptable and lead to desirable results.

\section{Simulation Studies} \label{ch3_EB_simulation}
\subsection{Simulation Settings}
We evaluated the performance of our proposed estimators through simulation studies in various settings and compared it to MLE from the direct regression, CSPML (denote as CML for simplicity) estimators and individual EB estimators, which incorporate single external model information. In each simulation, we compared six methods (direct regression, CML, EB, IVW, OCWE and SC-Learner) considering both overall and covariate-wise performance. We used the estimated standard error (ESE) to assess the precision gain of point estimates compared with the direct regression, and evaluated three overall metrics on a validation dataset of size $\rm N_{test}$=1,000:
\begin{itemize}
    \item Average estimated variance of logit-transformed predicted probability: \\$\rm \bar{V}[logit(\hat{p})]=$
    $\rm \frac{1}{N_{test}}\sum_{i=1}^{N_{test}}\hat{V}ar[logit(\hat{p}_i)]$, where $\rm \hat{p}$ denotes the estimated probability and $\rm logit(p) = log(\frac{p}{1-p})=(\rm \boldsymbol{\rm X},B) \boldsymbol{\gamma}$;
    \item Sum of squared errors: $\rm SSE=\frac{1}{N_{test}}\sum_{i=1}^{N_{test}}(\hat{p}_i-p_{i0})^2$, where $\rm \hat{p}_{i0}$ denotes the true probability of $\rm Y_i=1$ given $\rm X_i$ and $\rm B_i$;
    \item Scaled Brier score: $\rm \frac{1}{N_{test}}\sum_{i=1}^{N_{test}}(Y_i-\hat{p}_i)^2 / \frac{1}{N_{test}}\sum_{i=1}^{N_{test}}(Y_i-\bar{Y})^2$.
\end{itemize}

A summary of all the simulation settings is listed in Figure \ref{sim_setup}. In the first four simulation scenarios (I, II, III, IV), we assumed that a logistic regression model with the following form could describe the relationship among a binary outcome Y and five covariates ($\rm X_1,X_2,X_3,X_4, B$), where $\rm X_1,X_2,X_3,X_4$ had been used in at least one external model while $\rm B$ was only available in the current study:
$\rm logit[Pr(Y=1|X_1,X_2,X_3,X_4,B)] = -1 - 0.5\sum_{i=1}^4 X_i + 0.5B$.
Here $\rm X_1,X_2,X_3,X_4$ and $\rm B$ followed a standard multivariate normal distribution with 0.3 correlation and the prevalence of $\rm Y=1$ was 0.32. 
\begin{itemize}
    \item Simulation I evaluated the idealized case where the internal and the external models were fitted on the homogeneous population.
    \item Simulation II assessed the performance of the proposed method when external model 1 had biased $\rm \hat{\boldsymbol\beta}$ estimates fitted from $\rm Y|X$, where we obtained the incorrect model estimates by fitting external model 1 on a small dataset of size 500.
    \item Simulation III aims to show the impact of heterogeneous covariate distribution (X, B) in the external population. As \cite{Estes2017} assessed, the disparity of the (X, B) distribution between the two populations can come from (i) a different conditional distribution $\rm B|X$, (ii) a different marginal B distribution, (iii) a different marginal X distribution, or a combination of these reasons. In this simulation scenario, we showed the combination of (i) and (ii) as an example, but we will discuss the result of other scenarios in Simulation Results in Section 3.2.
    \item Simulation IV evaluated the situation where the outcome model was misspecified in external model 3.
\end{itemize}
We assumed that there was an internal study of size n=200 and three external models have been fitted to a very large synthetic dataset (sample size $\rm m_1= m_2= m_3=30,000$ for simplicity) that is sampled from the true data generating mechanism and gives precise estimates of the model parameters (except external model 1 in Simulation II). The external sample size need not be as large as 30,000 to achieve good performance as long as the estimated model parameters are close to the true parameters. Sensitivity analysis (results not shown) using external sample size $\rm m_1= m_2= m_3=1,000$ showed small numerical differences compared with m1= m2= m3=30,000. 

In simulation V and VI, we considered the outcome model with higher dimension and homogeneous populations between the internal and the external models. In these two scenarios, the internal sample size n=500 was used.
\begin{itemize}
    \item Simulation V evaluated the situation where the outcome model contained nine X's and one B: $\rm logit[Pr(Y=1|X_1,...,X_9,B)] = -1 - 0.5\sum_{i=1}^9 X_i + 0.5B$. Specifically, the external model 1 only contained two predictors, $\rm X_1$ and $\rm X_2$, the external model 2 contained seven predictors, $\rm X_1,...,X_7$, and the external model 3 contained six predictors $\rm X_1, X_2, X_3, X_4, X_7$ and $\rm X_8$. 
    \item simulation VI evaluated the situation where the full model contained three X's and five B's: $\rm logit[Pr(Y=1|X_1,X_2,X_3,B_1,...,B_5)] = -1 - 0.5\sum_{i=1}^3 X_i + 0.5\sum_{i=1}^5 B_i$,
using the same external models as simulation I.
\end{itemize}

\subsection{Simulation Results}
In Table \ref{Table_sim_1234} simulation I, we see that all CML and EB estimators are unbiased as expected. The estimated standard error (ESE, in square brackets) accurately reflect the true standard deviation (SD, in round brackets) from 500 simulations. Both OCWE and SC-Learner had better overall performance than single EB estimators, while SC-Learner outperformed OCWE with respect to both the covariate-wise and the overall metrics (Figure \ref{sim_plot_1234}).

In Table \ref{Table_sim_1234} simulation II, $\rm CML_1$ was biased with poor 95\% coverage rate due to biased estimation of $\boldsymbol{\hat{\beta}}$. Although trading off most of the precision gain, $\rm EB_1$ corrected the bias in $\rm CML_1$. The bias of $\rm CML_1$ also caused underestimation of the standard error of $\rm EB_1$ as highlighted in yellow, which was because $\rm \hat{\boldsymbol{\gamma}}_{CML_1}$ differed from $\rm \hat{\boldsymbol{\gamma}_I}$ as pointed out in Section \ref{Asymptotic_theory}. Despite that, both OCWE and SC-Learner detected the least efficiency gain of $\rm EB_3$ among all EB estimators and provided unbiased overall estimates. Moreover, OCWE and SC-Learner had the largest relative efficiency gain compared with the direct regression than other methods, as well as the best performance regarding the relative SSE and Brier Score gain only second to $\rm EB_3$ (Figure \ref{sim_plot_1234}).

Results of simulation III and IV in Table \ref{Table_sim_1234} indicates that when incorporating discordant external information from heterogeneous populations (one with different distribution of the joint [X,B] and another with different outcome model), $\rm EB$ estimators are able to correct the bias of $\rm CML$ estimators by trading off some precision-gain, and the proposed estimators could further identify and down-weight that particular $\rm EB$ estimators. \cite{Estes2017} provided comprehensive simulation results to show that the EB estimator can protect against the bias due to heterogeneous (X, B) distribution from the external model. Their results indicated that CML estimator would have a substantial bias when the difference came from the conditional distribution $\rm B|X$ or marginal distribution B, and marginal distribution X when the X-B interaction term is involved in the true $\rm Y|X$, B model. When evaluating these scenarios (summary and results in Appendix D in the Supplementary material), we found a similar pattern and conclusion as simulation III. In summary, these simulation results provided the evidence that the proposed approach is robust to the heterogeneous covariate distribution (X, B) of the external population. 

In addition to simulation IV, we have assessed different forms of misspecified outcome model in the external population listed in Table D3 from Appendix D of the Supplementary Material, including different intercept only and different X coefficients only, which showed similar conclusions. Simulation IV also showed that although higher dimension of the external model will lead to better overall prediction when the internal and external has the same population (simulation I and II), it is not the case when the transportability assumption is violated: OCWE identified that the external model 3 came from a different distribution than the internal study and thus assigned the smallest weight to $\rm EB_3$, even though it had the greatest number of predictors compared with the other two external models. In both simulations, OCWE and SC-Learner had similar covariate-wise and overall performances (Figure \ref{sim_plot_1234}).

Table \ref{Table_sim_56} further shows that the proposed estimators have decent performance when the number of predictors in the external models has large differences (simulation V) and when the dimension of B is larger than X's (simulation VI). In simulation V, compared with other external models that included more than six predictors, the external model 1 only used two predictors and thus provided the least amount of extra information. We see that OCWE assigns the minimum weight to $\rm EB_1$ and achieves the largest relative efficiency gain among all estimators (Figure \ref{sim_plot_56}). The fact that SC-Learner outperformed OCWE with respect to decreasing the covariate-wise variance of $\rm \hat{\gamma}_5$ to $\rm \hat{\gamma}_9$ reveals that SC-Learner is flexible enough to select the external information on the covariate-level when $\rm p_k$ (the number of predictors used in the $\rm k^{th}$ external model) is very different from others. Simulation VI showed that when the dimension of B was much larger than X, the overall benefit of combining multiple EB estimators would be limited due to the small amount of external information added from the external models.

\section{Application to prostate cancer data}
To assess the performance of the proposed estimators in a real data example, we developed a model for predicting the risk of high-grade prostate cancer (Gleason score $>$ 6) using a combination of a set of internal individual-level data and two external risk calculators from different studies. The first risk calculator was developed based on the Prostate Cancer Prevention Trial (PCPT) in the United States \citep{Thompson2006}. This calculator, denoted as PCPThg, is built on five clinical variables including prostate-specific antigen (PSA) level, digital rectal examination (DRE) findings, age, race (African American or not) and prior biopsy results using the following model:
\begin{equation}
    \rm logit(p_i)=-6.25+1.29log(PSA_i)+DRE_i+0.03Age_i+0.96Race_i-0.36Biopsy_i,
\label{PCPThg}
\end{equation}
where $\rm p_i$ is the probability of observing high-grade prostate cancer for subject i. The second risk calculator is the European Randomized Study of Screening for Prostate Cancer (ERSPC) risk calculator 3 \citep{ERSPC2012}, which uses slightly different clinical variables to predict the same risk as PCPThg: PSA level, DRE findings, and transrectal ultrasound prostate volume (TRUS-PV) in a logistic regression model shown as below:
\begin{equation}
    \rm logit(p_i)=-3.15+1.18log_2(PSA_i)+1.81DRE_i-1.51log_2(\text{TRUS-PV}_i),
\label{ERSPC}
\end{equation}
where TRUS-PV was categorized as a 3-category variable described in \cite{ERSPC2012}. In addition to all the predictors used in the external model \ref{PCPThg} and \ref{ERSPC}, we considered adding two more log-2-transformed biomarkers that had not been widely used but shown to be predictive of prostate cancer \citep{Tomlins2015, Truong2013}, prostate cancer antigen 3 (PCA3) and TMPRSS2:ERG (T2:ERG) gene fusions to our target model:
\begin{align*}
    \rm logit(p_i)= \rm \gamma_0 + &\gamma_1 \rm log_2(PSA_i) +\gamma_2 DRE_i +\gamma_3 Age_i + \gamma_4 Biopsy_i +\gamma_5 Race_i \\
    \rm +& \rm \gamma_6 log_2(\text{TRUS-PV}_i) +\gamma_7 log_2(PCA3_i+1)+ \gamma_8 log_2(\text{T2:ERG}_i+1),
\numberthis \label{realdata_full_model}
\end{align*}
Using the data from \cite{Tomlins2015}, we had 678 male patients in the internal dataset who had complete data of all eight covariates listed in model \ref{realdata_full_model} and an independent validation data of size 1,174 (sample size reduced from the initial 679 and 1218 patients, respectively due to missing TRUS-PV that was not previously used in \cite{Tomlins2015}). Details of the individual-level data, including the description of the internal and the validation dataset, and the recent applications using the same setting can be found in \cite{Tomlins2015} and \cite{Cheng2019}.

PCPThg risk calculator utilizes standard clinical and demographic variables that have been widely used while ERSPC additionally considers the prostate volume that was shown to be related to PSA level \citep{Bohnen2007}. In addition, similarities of prostate-specific antigen patterns between United States and European populations prostate-specific antigen patterns has been shown \citep{Simpkin2016}. Therefore, incorporating information from both risk calculators can potentially provide more accurate estimation of the risk parameters and narrower confidence bands, which could in turn yield better prediction performance and improved inference. 

In order to reconcile the discrepancy between the covariates used in the external models and have a compatible interpretation of the intercept, we centered all variables in the original models \ref{PCPThg} and \ref{ERSPC}, and log2-transformed PSA and TRUS-PV by adjusting the corresponding intercepts a-priori (details in Appendix E of the Supplementary Material). We present the estimated coefficients and standard errors in Table \ref{Table_realdata}. Similar to the simulation study, we calculated the scaled Brier Score and the average estimated variance of logit-transformed predicted probability based on the validation dataset as the prediction metrics. 

As shown in Table \ref{Table_realdata}, OCWE assigns almost zero weight to the ERSPC risk calculator, which indicated a large population discrepancy between the internal data and the underlying European population possibly due to the difference in the intercept, which reflects that the prevalence of high-grade prostate cancer is higher in the European population compared with patients in the United States who had average covariate values. Even though, SC-Learner was able to make the most of the little improvement provided by ERSPC risk calculator and augment the point-wise precision gain for covariate PSA, DRE and TRUS-PV (3\%, 4\% and 12\% more compared with OCWE, respectively), which led to the largest overall improvement as well (17.2 \% decrease of the average prediction variance compared with the direct regression).

\section{Discussion}
The proposed framework along with two weighted estimators, OCWE and SC-Learner, adds to the evolving research on using external summary-level information to bolster the statistical efficiency of the internal study for improved inference. This new method is flexible and robust in the ways that (i) it is capable of incorporating external models that use a slightly different set of covariates; (ii) it is able to identify the most relevant external information and diminish the influence of information that is less compatible with the internal data; and (iii) it nicely balances the bias-variance trade off while preserving the most precision gain. Moreover, our extensive simulation studies and the real data example shows that the proposed estimators are more efficient and robust than the naïve analysis of the internal data and other naïve combinations of external estimators in both idealized and non-idealized settings. 

Compared with a single EB estimator, the proposed composite estimators can have up to 32\% more improvement in MSE regarding one covariate (Figure \ref{sim_plot_1234}) and decent improvement regarding the overall metric such as 20\% further decrease in SSE and 11.5\% further decrease in estimated prediction variance (Figure \ref{sim_plot_1234}). In some cases, several single EB estimators that showed limited gains mitigate another single EB estimator’s excellent performance during integration, e.g. Simulation II. In reality, the proposed composite estimators will be preferred over a single EB estimator, since it is often difficult to pick the best external model that contains the most useful information to boost the inference of the internal study among several available external models.

In practice, the choice of SC-Learner or OCWE mainly depends on the features of the external models and the user’s research goal. As shown in simulation V, if at least one of the external models used very few predictors compared to the full dimension of (X, B), i.e., $\rm p_k << p$, we suggest using SC-Learner as it can adapt the external information being considered covariate-wise and thus prevent the gain in certain covariates from being washed away when the dimension of predictors are uneven. Similarly, we would recommend using SC-Learner if the researcher cares about maximizing the precision gain on the covariate-level or improving precision in certain covariates are of particular interest. On the other hand, if the research goal involves ranking the usefulness/relevancy of the external model, OCWE would be a good choice as it provides one unified weight for all covariates in the same model and outperforms comparable estimators such as IVW (inverse variance-weighed estimator).

A different approach to estimating the weights in OCWE would be cross-validation. For example, instead of minimizing the estimated prediction variance over all internal observations, one could randomly split the data into training and testing data, and choose the weights that minimize the objective function over the testing data only, then average this over different data splits. This approach could potentially prevent overfitting, give more stable estimates of the weights and improve the predictive performance.

As is typical of shrinkage estimators, in finite-sized samples, the EB method sacrifices a small amount of efficiency when the assumption of full transportability is satisfied, but reduces the potential bias of the CSPML when full transportability is not satisfied, while still being more efficient than the simple estimate from the internal data. In the proposed method, the magnitude of the precision-gain depends on the degree of the distributional similarity between the internal and external populations, i.e., the more similar, the more benefit we will gain by incorporating the external models. In the extreme case when these populations are completely different, our approach is very similar to analyzing the internal data only. On the contrary, when these populations share the identical joint distribution of (Y, X, B), we will achieve near to the maximum possible benefit.

Note that the proposed framework is not suitable if some predictors used in the external models are completely unmeasured in the internal study. In addition, the proposed framework is constructed based on parametric regression models, which requires the exact form of the external models and common covariates shared across different external models. In some cases, such as the real data example in this study, the authors were able to reconcile the discrepant transformations (i.e. one used natural-log PSA and the other used mean-centered log-2 based PSA) by reparametrizing the intercept. But this may not always be feasible, in which case we suggest considering methods that only require the predicted probability or the ability to estimate the probability given predictors without knowing the exact models, such as the synthetic data method proposed in \cite{Gu2019}. It is plausible that the assumed parametric model is not a good approximation of the internal data’s underlying distribution. In the simulation (results not shown here), we saw that when the effect of non-linear terms was small, the proposed method could still correctly estimate the main effect. Besides, some reassurance about the selected structure of the parametric internal model can be obtained from the external models. The external models determine the X variables that are included and how they are included. For example, in our prostate cancer example, the external models took a log transformation of PSA, thus the parametric model for the internal data also includes a log transformation of PSA. Since the external datasets are typically large, we might surmise that if a large non-linearity or a strong interaction amongst the X variables were needed, it would have been included in the external model.

Recently, \citet{Zhang2020} proposed a general framework as the extension of \citet{Chatterjee2016} to solve the same genre of problem when the external parameter uncertainty cannot be ignored. In the situation where the external study population differs from the internal one, the performance of their method depends on the availability of high-quality reference data from the external population, similar to in \citet{Chatterjee2016}. Our method also provides the option of incorporating external parameter uncertainty but more importantly it provides valid internal inference when it is in general hard to obtain the right reference data in reality. In addition, \citet{Chatterjee2019} proposed a generalized meta-analysis framework building from the  \citet{Chatterjee2016} approach to combine information of multiple regression models with disparate covariates using a method of moment approach. Different from our goal, their method is an extension of the fixed-effect meta-analysis that also relies on the existence of reference data, and the performance of the proposed estimator depends on the quality and representativeness of the reference data. 

One possible extension of the proposed method is the application in causal inference. If the treatment indicator was available as one of the X covariates, one could directly calculate the estimated average causal effect through the formula $\rm E_{X,B}[E(Y| treatment=1, X, B)-E(Y| treatment=0, X, B)]$ using the regression estimates obtained from the proposed method. \cite{YangDing2020} considered a similar setting, where they view the internal data as the validation data with richer covariates while the external data serves as the main dataset with fewer covariates, aiming to improve the efficiency of the initial estimator $\rm \hat{\boldsymbol\gamma}_I$ from the internal dataset by incorporating a constructed zero-mean error-prone estimator $\rm \hat{\boldsymbol\beta}_I-\hat{\boldsymbol\beta}_E$, where $\rm \hat{\boldsymbol\beta}_I$ and $\rm \hat{\boldsymbol\beta}_E$ are the estimators using X only from the internal and external population, respectively. Using notation from \cite{Estes2017}, we can also reparametrize Yang and Ding’s estimator as a weighted average of $\rm \hat{\boldsymbol\gamma}_I$ and $\rm \hat{\boldsymbol\gamma}_{CML}$, where the only difference is that the EB estimator has the shrinkage effect by empirically estimating the variance-covariance matrix that plays an important role in the weights.

It is worth noting that there is a popular field in machine learning called ensemble learning with a large and evolving literature, aiming to combine several base models to produce the optimal predictive model. Some representative ensemble methods include but not limited to Boosting \citep{Schapire1990}, Bagging \citep{Freund1997} and Stacking \citep{Breiman1996} with some examples being random forest \citep{Breiman2001} and Super Learner \citep{VanderLaan2007}. The key difference of our proposed method is that we have a specific parametric model of interest, and we are taking the weighted average of the estimated coefficients of that model from several estimators such that we can measure the impact of each predictor and its uncertainty, instead of directly weighting the predicted outcomes as in these ensemble methods. The proposed method can provide competitive and robust estimators for statistical inference. Moreover, the proposed estimators have improved efficiency compared with direct regression using the internal data only or naïve inverse variance-weighted estimator. However, if the research goal is to find the optimal predictive model with the minimum prediction error solely, especially when the underlying mechanism is not of interest, it would be worthwhile to explore the field of ensemble methods mentioned above, which is beyond the discussion of this study.

Last but not least, the issue of transportability of risk prediction models is a critical one and one that is often encountered in practice. While it is plausible that the association between pairs of variables or even the joint distribution of all the variables is similar between populations, it is also plausible that they could differ, not just due to biological or behavioral differences in populations but also due to being collected in different parts of the world or different decades.  The EB strategy will be a good choice, balancing between bias and efficiency when one is unsure about whether transportability assumptions hold for risk models across time, space or cohorts.

\section{SUPPLEMENTARY MATERIAL}
A web-based Supplementary Material is available online. R package MetaIntegration is available at https://github.com/umich-biostatistics/MetaIntegration.

\section*{ACKNOWLEDGEMENTS}
This research was partially supported by National Institutes of Health grant CA129102 and NSF DMS grant 171933.

\bibliographystyle{biorefs}
\bibliography{refs}


\begin{landscape}

\begin{figure}[H]
\centering
\includegraphics[width=0.88\linewidth] {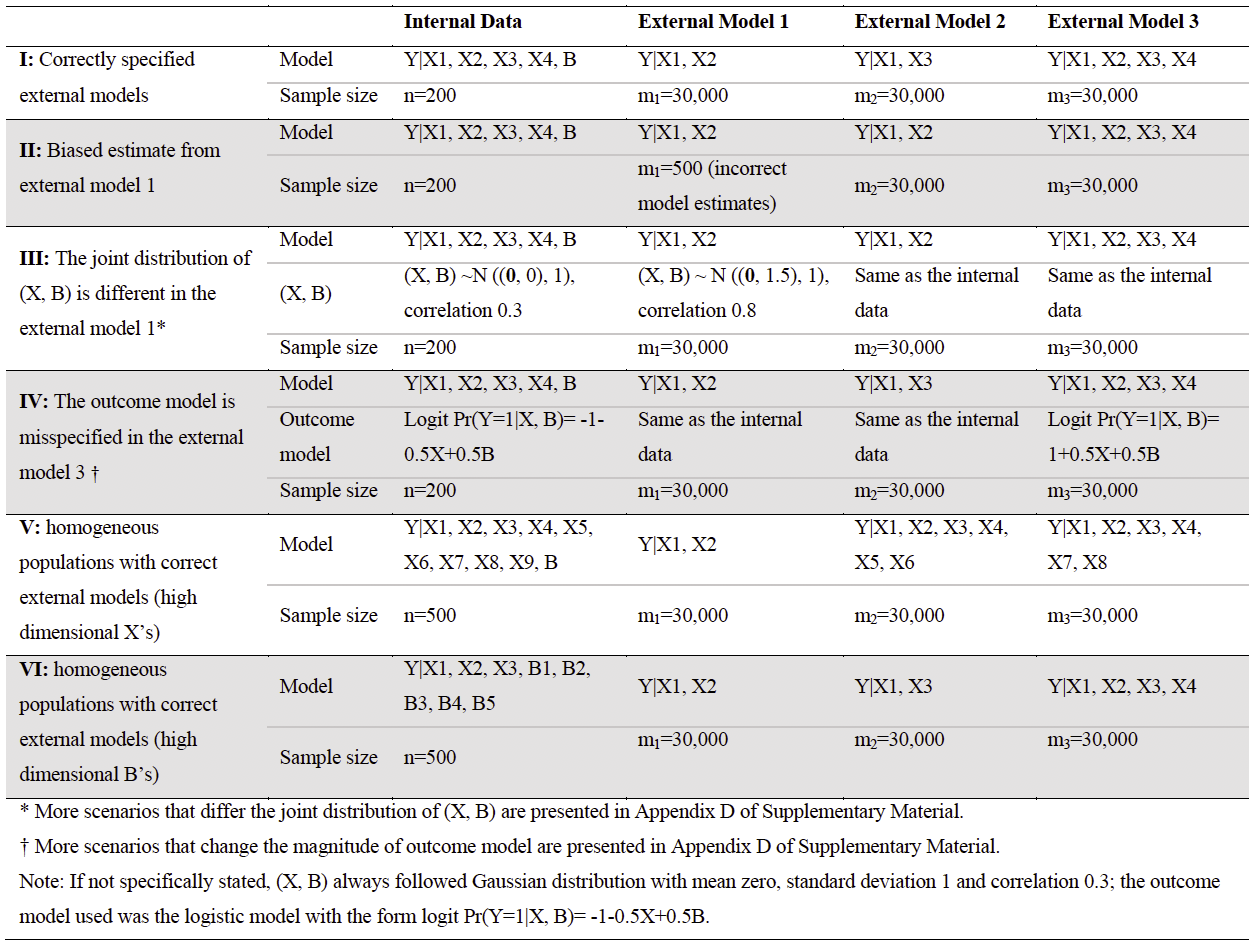}
\caption{Simulation Settings Snapshot} \label{sim_setup}
\end{figure}

\begin{figure}[H]
\adjustimage{width=1.2\linewidth, center=1.1\linewidth}{sim_plot_1234}
\caption{Visualization of the metrics to evaluate the performance in simulations I-IV. Scatter plot shows the covariate-wise relative mean square error (MSE) improvement compared with the direct regression (x axis represents $\rm \hat{\gamma}_X$; $\rm \hat{\gamma}_B$ not shown since no external information incorporated; larger y values represent larger MSE gain) while the line plots represent the relative efficiency/SSE/BS gain compared with the direct regression fitting on a validation dataset of size 1,000 (longer lines represent larger gain). Abbreviations: SSE, sum of square error; Direct, direct regression; EB, empirical Bayes method; SD, Monte Carlo standard deviation from 500 simulations; ESE, estimated standard error; IVW, inverse variance-weighted estimator; OCWE, optimal covariance-weighted estimator; SC-Learner, selective coefficient learner.} \label{sim_plot_1234}
\end{figure}

\begin{figure}[H]
\adjustimage{width=1.1\linewidth, center=1.1\linewidth}{sim_plot_56}
\caption{Visualization of the metrics to evaluate the performance in simulations V-VI. Scatter plot shows the covariate-wise relative mean square error (MSE) improvement compared with the direct regression (x axis represents $\rm \hat{\gamma}_X$; $\rm \hat{\gamma}_B$ not shown since no external information incorporated; larger y values represent larger MSE gain) while the line plots represent the relative efficiency/SSE/BS gain compared with the direct regression fitting on a validation dataset of size 1,000 (longer lines represent larger gain). Abbreviations: SSE, sum of square error; Direct, direct regression; EB, empirical Bayes method; SD, Monte Carlo standard deviation from 500 simulations; ESE, estimated standard error; IVW, inverse variance-weighted estimator; OCWE, optimal covariance-weighted estimator; SC-Learner, selective coefficient learner.} \label{sim_plot_56}
\end{figure}

\end{landscape}

\begin{sidewaystable}
\caption{\it{Results of simulation I--IV. Internal dataset had size n=200; green represents good performance with small bias and large efficiency gain, yellow represents underestimated ESE (in square brackets) compared with the Monte Carlo SD (in round brackets), and red represents poor performance of bias/95\% coverage rate.} \label{Table_sim_1234}}
\begin{adjustbox}{max width=1\linewidth, center=1\linewidth, max height=1\linewidth}
\arrayrulecolor{black}
\begin{tabular}{llllllllllll} 
\hline
\multicolumn{2}{l}{\multirow{4}{*}{Simulation}} & \multirow{2}{*}{Direct Regression} & \multicolumn{2}{l}{Internal Data + External 1} & \multicolumn{2}{l}{Internal Data + External 2} & \multicolumn{2}{l}{Internal Data + External 3} & \multicolumn{3}{l}{Composite of EB Estimators}  \\ 
\cline{4-12}
~ & ~ & ~ & CML 1 & EB 1  & CML 2 & EB 2  & CML 3 & EB 3  & IVW & OCWE & SC-Learner  \\ 
\cline{3-12}
~ & ~ &\multicolumn{10}{l}{Bias (SD) [ESE]}                        \\
~ & ~ & \multicolumn{10}{l}{95\% Coverage Rate}                    \\
\hline
\multirow{13}{*}{I} & Weight & / & / & / & / & / & / & / & (.33, .33,.33) & {\cellcolor[rgb]{0.886,0.937,0.851}}{(}.27, .12, .61) & /
\\
~ & \multirow{2}{*}{$\rm \gamma_0$} 
& -.041 (.197) [.193] & -.008 (.081) [.085]                            
& -.032 (.162) [.160]                                     
& -.007 (.089) [.084]                                     
& -.032 (.163) [.161]                                     
& .004 (.050) [.085]                    
& -.033 (.169) [.168]                   
& {\cellcolor[rgb]{0.886,0.937,0.851}}-.032 (.164) [.162] 
& {\cellcolor[rgb]{0.886,0.937,0.851}}-.031 (.166) [.164] 
& {\cellcolor[rgb]{0.886,0.937,0.851}}-.032 (.164) [.162]  
\\
~ & ~
& 95\%                
& 96\%                                                    
& 95\%                                                    
& 95\%                                                    
& 95\%                                                    
& 98\%                                  
& 95\%                                  
& {\cellcolor[rgb]{0.886,0.937,0.851}}95\%                
& {\cellcolor[rgb]{0.886,0.937,0.851}}95\%                
& {\cellcolor[rgb]{0.886,0.937,0.851}}95\%                
\\
\hline
\end{tabular}
\arrayrulecolor{black}
\end{adjustbox}
\end{sidewaystable}

\begin{sidewaystable}
\renewcommand\thetable{1 (Cont'd)}
\caption{\it{Results of simulation V--VI. Internal dataset had size n=500; green represents good performance with small bias and large efficiency gain.} \label{Table_sim_56}}
\begin{adjustbox}{max width=1\linewidth, center=1\linewidth, max height=1\linewidth}
\arrayrulecolor{black}
\begin{tabular}{llllllllllll} 
\hline
\multicolumn{2}{l}{\multirow{4}{*}{Simulation}} & \multirow{2}{*}{Direct Regression} & \multicolumn{2}{l}{Internal Data + External 1} & \multicolumn{2}{l}{Internal Data + External 2} & \multicolumn{2}{l}{Internal Data + External 3} & \multicolumn{3}{l}{Composite of EB Estimators}  \\ 
\cline{4-12}
~ & ~ & ~ & CML 1 & EB 1  & CML 2 & EB 2  & CML 3 & EB 3  & IVW & OCWE & SC-Learner  \\ 
\cline{3-12}
~ & ~ &\multicolumn{10}{l}{Bias (SD) [ESE]}                        \\
~ & ~ & \multicolumn{10}{l}{95\% Coverage Rate}                    \\
\hline
\multirow{23}{*}{V} & Weight & / & / & / & / & / & / & / & (.33, .33,.33) & {\cellcolor[rgb]{0.886,0.937,0.851}}{(.13, .432, .429)} & /
\\
~ & \multirow{2}{*}{$\rm \gamma_0$} 
& -.035 (.153) [.149] 
& .02 (.100) [.098]
& .031 (.132) [.129]                                      
& -.002 (.054) [.059]                                     
& -.031 (.139) [.137]                                     
& -.010 (.064) [.066]                                     
& -.031 (.137) [.135]                                     
& {\cellcolor[rgb]{0.886,0.937,0.851}}-.031 (.135) [.132] 
& {\cellcolor[rgb]{0.886,0.937,0.851}}-.031 (.137) [.133] 
& {\cellcolor[rgb]{0.886,0.937,0.851}}-.031 (.135) [.132]  
\\
~ & 
& 94\%                
& 95\%                                                    
& 95\%                                                    
& 97\%                                                    
& 94\%                                                    
& 96\%                                                    
& 94\%                                                    
& {\cellcolor[rgb]{0.886,0.937,0.851}}94\%                
& {\cellcolor[rgb]{0.886,0.937,0.851}}94\%                
& {\cellcolor[rgb]{0.886,0.937,0.851}}94\%                 
\\
\hline
\end{tabular}
\arrayrulecolor{black}
\end{adjustbox}
\end{sidewaystable}

\begin{sidewaystable}
\renewcommand\thetable{2}
\caption{\it{Results of the real data example for predicting risk of high-grade prostate cancer. Internal dataset of size n=678; validation dataset of size $\rm N_{test}$=1,174; ESE, estimated standard error using the asymptotic formula;
$\downarrow \%$, percentage of ESE decrease compared with the direct regression; PCPThg, prostate cancer prevention trial risk calculator; ERSPC, the European Randomized study of Screening for Prostate Cancer risk calculator 3; IVW, inverse variance-weighted estimators; OCWE, optimal covariate weighted estimator; SC-Learner, selective coefficient learner; REF, reference; green represents good performance with small bias and large efficiency gain, and grey represents no efficiency improvement due to covariate not included in the external calculator.} 
\label{Table_realdata}}
\begin{adjustbox}{max width=1\linewidth, center=1\linewidth, max height=1\linewidth}
\arrayrulecolor{black}
{\tabcolsep=4.25pt
\begin{tabular}{@{}lllllllllllll@{}}
\hline
\multirow{2}{*}{~}  & 
\multirow{2}{*}{Direct Regression} & 
\multicolumn{2}{l}{PCPThg} & 
\multicolumn{2}{l}{ERSPC} & 
\multicolumn{2}{l}{Internal data + PCPThg} & 
\multicolumn{2}{l}{Internal data + ERSPC}  & 
\multicolumn{3}{l}{Composite of EB Estimators}
\\ 
\cline{3-13}& ~ & Original & Estimated & Original & Estimated & CML 1 & EB1 & CML 2  & EB2 & IVW  & OCWE & SC-Learner
\\ 
\hline
Weight  & / & / & / & / & / & / & / & / & / & {[}.5, .5] & {\cellcolor[rgb]{0.886,0.937,0.851}}{[}.99, .00]   & / 
\\
$\rm \bar{V}[logit(\hat{p})]$  & .1199 & / & .065 & / & .048 & .073  & .104  & .078  & .105 & .104 ($\downarrow$ 13.2\%) & .104 ($\downarrow$ 13.2\%) & {\cellcolor[rgb]{0.886,0.937,0.851}}.0996 ($\downarrow$ 17.2\%)  
\\
Brier Score & .902 
& 1.059  
& .987   
& .954  
& .994   
& .942   
& .901   
& .847   
& .901   
& {\cellcolor[rgb]{0.886,0.937,0.851}}.901 ($\downarrow$ .12\%) 
& {\cellcolor[rgb]{0.886,0.937,0.851}}.901 ($\downarrow$ .12\%) 
& {\cellcolor[rgb]{0.886,0.937,0.851}}.901 ($\downarrow$ .12\%)
\\ 
\hline
~ & \multicolumn{5}{l}{Point estimate (ESE)} 
\\
~ & \multicolumn{7}{l}{$\downarrow$\% ESE w.r.t direct regression} 
\\
\hline
\multirow{2}{*}{Intercept}  & -4.123 (.449)  & -3.686268 & -1.395 (.115) & -3.16 & -1.382 (.111) & -6.422 (.437)  & -4.128 (.445) & -5.989 (.437) & -4.130 (.444) & {\cellcolor[rgb]{0.886,0.937,0.851}}-4.129 (.443) & {\cellcolor[rgb]{0.886,0.937,0.851}}-4.128 (.445)  & {\cellcolor[rgb]{0.886,0.937,0.851}}-4.129 (.443)  
\\
& REF   
& / 
&  74\% 
& / 
&  75\% 
&  3\% 
&  1\% 
&  3\% 
&  1\% 
& {\cellcolor[rgb]{0.886,0.937,0.851}}  1\%   
& {\cellcolor[rgb]{0.886,0.937,0.851}}  1\%   
& {\cellcolor[rgb]{0.886,0.937,0.851}}  1\% 
\\
\multirow{2}{*}{$\rm log_2$(PSA)} 
& 0.860 (.144) 
& 0.8941599 
& 0.721 (.123) 
& 1.175573 
& .878 (.123) 
& 0.893 (.103) 
& 0.860 (.132) 
& 1.159 (.081) 
& 0.861 (.124) 
& {\cellcolor[rgb]{0.886,0.937,0.851}}0.860 (.127)  
& 0.860 (.130) & {\cellcolor[rgb]{0.886,0.937,0.851}}0.860 (.127)
\\
& REF   
& /  
&  14\%  
& /  
&  14\% 
&  28\%  
&  8\%  
&  44\% 
&  14\% 
& {\cellcolor[rgb]{0.886,0.937,0.851}} 11\% 
&  9\% 
& {\cellcolor[rgb]{0.886,0.937,0.851}} 12\%  
\\
\multirow{2}{*}{DRE}  
& 1.028 (.297)  
& 1 & 1.134 (.256)  
& 1.813195  
& 1.298 (.269) 
& 0.687 (.218)  
& 1.027 (.269)  
& 1.430 (.170)  
& 1.029 (.253) 
& {\cellcolor[rgb]{0.886,0.937,0.851}}1.028 (.259)  
& 1.027 (.269)  
& {\cellcolor[rgb]{0.886,0.937,0.851}}1.028 (.258)     
\\
& REF  
& / 
& 14\% 
& / 
&  9\%  
&  26\% 
&  9\%  
&  43\%  
&  15\%   
& {\cellcolor[rgb]{0.886,0.937,0.851}} 13\%  
&  9\%    
& {\cellcolor[rgb]{0.886,0.937,0.851}} 13\% 
\\
\multirow{2}{*}{Age} 
& 0.0315 (.014)     
& 0.03         
& 0.0328 (.012)  
& {\cellcolor[rgb]{0.906,0.902,0.902}} 
& {\cellcolor[rgb]{0.906,0.902,0.902}} 
& 0.0304 (.009)  
& 0.0315 (.013)  
& {\cellcolor[rgb]{0.906,0.902,0.902}}0.032 (.014)  
& {\cellcolor[rgb]{0.906,0.902,0.902}}0.031 (.014) 
& 0.032 (.014)   
& {\cellcolor[rgb]{0.886,0.937,0.851}}0.032 (.013)  
& {\cellcolor[rgb]{0.886,0.937,0.851}}0.032 (.013)     
\\
& REF   
& /   
& 14\% & \multirow{-2}{*}{{\cellcolor[rgb]{0.906,0.902,0.902}}~} 
& \multirow{-2}{*}{{\cellcolor[rgb]{0.906,0.902,0.902}}~} 
& 31\% & 8\% & {\cellcolor[rgb]{0.906,0.902,0.902}}~  
& {\cellcolor[rgb]{0.906,0.902,0.902}}~  
&  1\% & {\cellcolor[rgb]{0.886,0.937,0.851}} 8\%  
& {\cellcolor[rgb]{0.886,0.937,0.851}} 8\%  
\\
\multirow{2}{*}{Biopsy} 
& -1.165 (.290) 
& -0.36 
& -1.413 (.270)  
& {\cellcolor[rgb]{0.906,0.902,0.902}}                  
& {\cellcolor[rgb]{0.906,0.902,0.902}}                  
& -0.0286 (.158) & -1.163 (.255)  
& {\cellcolor[rgb]{0.906,0.902,0.902}}-1.187 (.290) 
& {\cellcolor[rgb]{0.906,0.902,0.902}}-1.165 (.291) 
& -1.164 (.272)  & {\cellcolor[rgb]{0.886,0.937,0.851}}-1.163 (.255)  
& {\cellcolor[rgb]{0.886,0.937,0.851}}-1.163 (.255)    
\\
& REF   
& /  
&  7\% 
& \multirow{-2}{*}{{\cellcolor[rgb]{0.906,0.902,0.902}}~} 
& \multirow{-2}{*}{{\cellcolor[rgb]{0.906,0.902,0.902}}~} 
&  46\%  
&  12\%   
& {\cellcolor[rgb]{0.906,0.902,0.902}}~  
& {\cellcolor[rgb]{0.906,0.902,0.902}}~  
&  6\%  & {\cellcolor[rgb]{0.886,0.937,0.851}} 12\%   
& {\cellcolor[rgb]{0.886,0.937,0.851}} 12\%   
\\
\multirow{2}{*}{Race} & 0.193 (.329)  & 0.96  & 0.448 (.287)  & {\cellcolor[rgb]{0.906,0.902,0.902}} & {\cellcolor[rgb]{0.906,0.902,0.902}}    & 0.819 (.218)  & 0.194 (.291)  & {\cellcolor[rgb]{0.906,0.902,0.902}}0.160 (.329)  & {\cellcolor[rgb]{0.906,0.902,0.902}}0.193 (.329)  & 0.193 (.308)   & {\cellcolor[rgb]{0.886,0.937,0.851}}0.194 (.291)   & {\cellcolor[rgb]{0.886,0.937,0.851}}0.194 (.291) 
\\
& REF   & /   & 13\%  & \multirow{-2}{*}{{\cellcolor[rgb]{0.906,0.902,0.902}}~} & \multirow{-2}{*}{{\cellcolor[rgb]{0.906,0.902,0.902}}~} &  34\%   &  12\%  & {\cellcolor[rgb]{0.906,0.902,0.902}}~ & {\cellcolor[rgb]{0.906,0.902,0.902}}~ &  6\%  & {\cellcolor[rgb]{0.886,0.937,0.851}} 12\%   & {\cellcolor[rgb]{0.886,0.937,0.851}} 12\%           
\\
\multirow{2}{*}{$\rm log_2$ (TRUS-PV)} & -1.663 (.252) & {\cellcolor[rgb]{0.906,0.902,0.902}~}                    & {\cellcolor[rgb]{0.906,0.902,0.902}~} & -1.514128 & -1.681 (.224)  & {\cellcolor[rgb]{0.906,0.902,0.902}}-1.683 (.252) & {\cellcolor[rgb]{0.906,0.902,0.902}}-1.663 (.252) & -1.491 (.155) & -1.663 (.223) & -1.663 (.236) & -1.663 (.252)  & {\cellcolor[rgb]{0.886,0.937,0.851}}-1.663 (.223)    
\\
& REF  & \multirow{-2}{*}{{\cellcolor[rgb]{0.906,0.902,0.902}}~} & \multirow{-2}{*}{{\cellcolor[rgb]{0.906,0.902,0.902}}~} & /  & 11\%  & {\cellcolor[rgb]{0.906,0.902,0.902}}  & {\cellcolor[rgb]{0.906,0.902,0.902}}  &  38\%  &  12\%  &  6\%  &  0\%  & {\cellcolor[rgb]{0.886,0.937,0.851}} 12\%           
\\
\multirow{2}{*}{$\rm log_2$ (PCA3+1)}    & .485 (.088)    & {\cellcolor[rgb]{0.906,0.902,0.902}}                  & {\cellcolor[rgb]{0.906,0.902,0.902}}                  & {\cellcolor[rgb]{0.906,0.902,0.902}}                  & {\cellcolor[rgb]{0.906,0.902,0.902}}                  & {\cellcolor[rgb]{0.906,0.902,0.902}}0.495 (.088)  & {\cellcolor[rgb]{0.906,0.902,0.902}}0.485 (.088)  & {\cellcolor[rgb]{0.906,0.902,0.902}}0.507 (.088)  & {\cellcolor[rgb]{0.906,0.902,0.902}}0.485 (.088)  & {\cellcolor[rgb]{0.906,0.902,0.902}}0.485 (.088)  & {\cellcolor[rgb]{0.906,0.902,0.902}}0.485 (.088)   & {\cellcolor[rgb]{0.906,0.902,0.902}}0.486 (.088)     
\\
& REF  & 
\multirow{-2}{*}{{\cellcolor[rgb]{0.906,0.902,0.902}}~} & 
\multirow{-2}{*}{{\cellcolor[rgb]{0.906,0.902,0.902}}~} & 
\multirow{-2}{*}{{\cellcolor[rgb]{0.906,0.902,0.902}}~} & \multirow{-2}{*}{{\cellcolor[rgb]{0.906,0.902,0.902}}~} & {\cellcolor[rgb]{0.906,0.902,0.902}}~  & {\cellcolor[rgb]{0.906,0.902,0.902}}~     & {\cellcolor[rgb]{0.906,0.902,0.902}}~       & {\cellcolor[rgb]{0.906,0.902,0.902}}~       & {\cellcolor[rgb]{0.906,0.902,0.902}}~       & {\cellcolor[rgb]{0.906,0.902,0.902}}~       & {\cellcolor[rgb]{0.906,0.902,0.902}}~                
\\
\multirow{2}{*}{$\rm log_2$ (T2:ERG+1)}  & .096 (.037)    & {\cellcolor[rgb]{0.906,0.902,0.902}}   & {\cellcolor[rgb]{0.906,0.902,0.902}}   & {\cellcolor[rgb]{0.906,0.902,0.902}}   & {\cellcolor[rgb]{0.906,0.902,0.902}}   & {\cellcolor[rgb]{0.906,0.902,0.902}}0.095 (.037)  & {\cellcolor[rgb]{0.906,0.902,0.902}}0.096 (.037)  & {\cellcolor[rgb]{0.906,0.902,0.902}}0.099 (.037)  & {\cellcolor[rgb]{0.906,0.902,0.902}}0.096 (.037)  & {\cellcolor[rgb]{0.906,0.902,0.902}}0.096 (.037)  & {\cellcolor[rgb]{0.906,0.902,0.902}}0.096 (.037)  & {\cellcolor[rgb]{0.906,0.902,0.902}}0.096 (.037)     
\\
& REF & 
\multirow{-2}{*}{{\cellcolor[rgb]{0.906,0.902,0.902}}~} & \multirow{-2}{*}{{\cellcolor[rgb]{0.906,0.902,0.902}}~} & \multirow{-2}{*}{{\cellcolor[rgb]{0.906,0.902,0.902}}~} & \multirow{-2}{*}{{\cellcolor[rgb]{0.906,0.902,0.902}}~} & {\cellcolor[rgb]{0.906,0.902,0.902}}~   & {\cellcolor[rgb]{0.906,0.902,0.902}}~             & {\cellcolor[rgb]{0.906,0.902,0.902}}~             & {\cellcolor[rgb]{0.906,0.902,0.902}}~             & {\cellcolor[rgb]{0.906,0.902,0.902}}~             & {\cellcolor[rgb]{0.906,0.902,0.902}}~              & {\cellcolor[rgb]{0.906,0.902,0.902}}~                
\\ 
\hline
\end{tabular}}
\arrayrulecolor{black}
\end{adjustbox}
\end{sidewaystable}

\end{document}